\documentclass[aps,prl,reprint,nofootinbib,superscriptaddress]{revtex4-1}

\usepackage{amssymb}
\usepackage{etoolbox}
\usepackage{color} 
\usepackage{tabularx}
\usepackage{subfigure}
\usepackage{amsmath,amssymb}
\usepackage{latexsym}
\usepackage{longtable}
\usepackage{psfrag}
\usepackage{epsfig}
\usepackage{graphicx,subfigure}
\usepackage{array}
\usepackage{tabularx}
\usepackage{hyperref}

\usepackage{multirow}
\usepackage{epstopdf}
\usepackage{graphicx}

\usepackage{algorithm}

\begin{document}
\title{Two-qubit state recovery from amplitude damping based on weak measurement}
\author{S. Harraz}
\affiliation{Department of Automation, University of Science and Technology of China, Hefei 230027, P. R. China}%
\author{S. Cong}
\email{scong@ustc.edu.cn}
\affiliation{Department of Automation, University of Science and Technology of China, Hefei 230027, P. R. China}%
\author{K. Li}
\email{kezhi.li@imperial.ac.uk}
\affiliation{Department of Electrical and Electronic Engineering, Imperial College London, SW7 2AZ, UK}

\begin{abstract}
In the quantum control process, arbitrary pure or mixed initial states need to be protected from amplitude damping through the noise channel using measurements and quantum control. However, how to achieve it on a two-qubit quantum system remains a challenge. In this paper, we propose a feed-forward control approach to protect arbitrary two-qubit pure or mixed initial states using the weak measurement.
A feed-forward operation and measurements are used before the noise channel, and afterwards a reversed operation and measurements are applied to recover the state back to its initial state. In the case of two-qubit pure states, we use the unravelling trick to describe the state of the system in each step of the control procedure. For two-qubit mixed states, a completely-positive trace-preserving (CPTP) map is implemented. Finally, the fidelity and success probability are used to evaluate the effect of protection. The complete recovery conditions for the measurement strengths are derived, under which we achieve the optimal fidelity and the success probability of recovering the initial pure or mixed states. 
\end{abstract}

\maketitle

\section{ I. introduction}\label{sec:sec1}

The dynamic of open quantum systems becomes decoherent easily due to the inevitable interaction with the environment \cite{lab1}. To suppress the effect of decoherence, various strategies have been studied, such as decoherence-free subspaces \cite{lab2}, quantum error correction \cite{lab3}, \cite{lab4}, quantum feedback control (QFBC) \citep{lab5,lab6,lab7} and quantum feed-forward control (QFFC) \citep{lab8,lab9,lab10}. For the problem of protecting quantum states from the noise, a quantum feedback control scheme was proposed \cite{lab11}, which included quantum weak measurement and correction rotation based on the result of measurements after the noise channel. It was experimentally implemented \cite{lab5}. This scheme was studied for different initial states, measurements and feedback control bases \citep{lab6,lab7}. In both quantum feedback control and quantum feed-forward control, the measurement plays an important role. Different from the measurement in the classical theory, in the quantum measurement there is a trade-off between the information gain and the disturbance of the system via measurement \cite{lab12}. Hence, the quantum weak measurement technique is promising, since it has little effects on the dynamic of the quantum system. Weak measurements generalise ordinary quantum measurements, and they reveal some information about the quantum state without collapsing the state into eigenvectors. This process is achieved by leveraging a weak coupling between the measurement device and the system.

Amplitude damping is a major decoherence that occurs in many quantum systems \cite{lab8}, such as a photon qubit in a leaky cavity, an atomic qubit subjected to spontaneous decay, or a super-conduction qubit with zero-temperature energy relaxation. The specific control problem we are interested in here is the stabilisation against amplitude damping for two-qubit pure or mixed states. Similar problem has been considered for one qubit quantum state recovery based on quantum feed-forward control. It is shown that one-qubit state affected by amplitude damping can be completely recovered by applying feed-forward control \cite{lab8}. Feed-forward control can be applied to make the state of the system immune to the effect of an amplitude damping channel. A feed-forward control (FFC) technique was used to realise a better impact of the discrimination of two nonorthogonal states after passing an amplitude damping channel \cite{lab13}. Also, a quantum composite control scheme was proposed \cite{lab14}, where quantum feedback control and quantum feed-forward control were combined for protecting two nonorthogonal states of a two-level quantum system against the amplitude damping noise. In \citep{lab9,lab10} the problem of protecting completely unknown states against given noise was mathematically discussed in two cases: only after-the-noise, and both before-and-after-the-noise. They mathematically shown that by using only after-the-noise control, one essentially cannot suppress any given noise. In other words, if the initial state is unknown, only the unitary rotational part of noise can be eliminated. Next, they considered the before-and-after-noise control scheme. It was shown that if the noise is weak, doing nothing to the system is the best control scheme; In case of intense noise, one needs to measure the system before noise and after the noise, based on the result of the measurement, reconstruct the state. Similar problems were considered for two-qubit quantum state recovery based on quantum gates \cite{lab15}. An arbitrary two-qubit pure state under amplitude damping in a weak measurement was probabilistically recovered using Hadamard and CNOT gates. However, their scheme cannot recover some states, in which they solved it by adding a step before the noise to prepare the system in a more robust state. Later, the authors used the same method to protect an arbitrary two-qubit mixed state \cite{lab16}.

In this paper, we consider the feed-forward control scheme for recovering arbitrary pure and mixed initial two-qubit states. We use the pre-weak measurement to gain information about the initial state. Then to make the states almost immune to the amplitude damping channel, we apply feed-forward operation based on the result of measurements. After the noise channel, we restore the initial state; hence, a reversed unitary operation and a post-weak measurement are applied. For mixed initial states, we propose a completely-positive trace-preserving (CPTP) map to describe the recovery control and final state of the system. we use the Monte-Carlo method over a large ensemble of initial states in experimental simulations to prove the effectiveness of feed-forward control for any arbitrary initial state. Furthermore, We compare the feed-forward control for two-qubits with the one in \cite{lab15}, and prove that feed-forward control has much better performance.

The paper's structure is as follows. In Sec. II. we introduce the feed-forward recovery control in general to protect arbitrary two-qubit states. 
In Sec. III, the protection of two-qubit pure initial states is specifically studied, and Sec. IV focuses on the feed-forward protection control in the case of mixed initial states. In Sec. V we give the complete recovery condition under which one can completely recover the state of the system.  In Sec. VI we show the behavior of the system in general protection schemes. Finally, the conclusion is drawn in Sec. VII.

\section{II. Weak-measurement based Feed-forward recovery control }\label{sec:sec2}
\begin{figure}[!htbp]
\centering
\includegraphics[height=7.5cm, width=4cm]{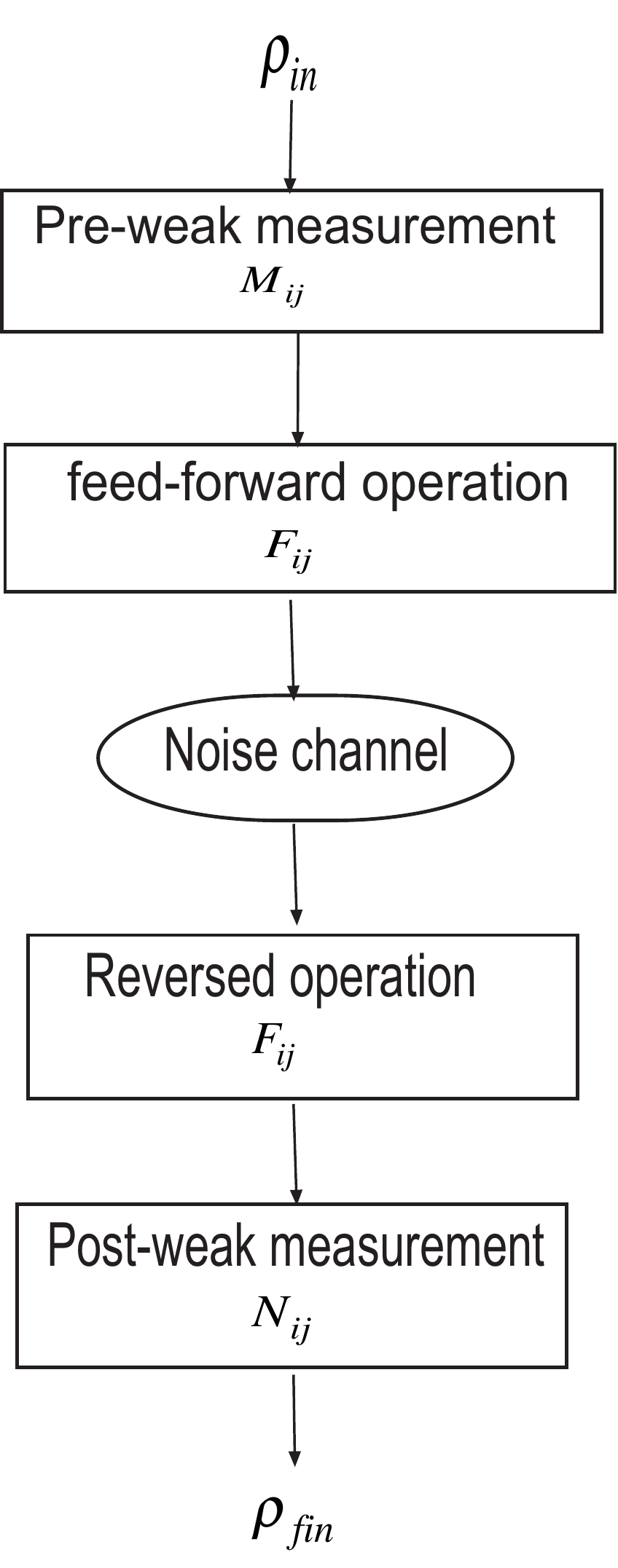}
\caption{The schematic diagram of the feed-forward recovery control scheme}
\label{fig1}
\end{figure}

The control task in this paper is to bring the state of the system after passing through the noise channel back to its initial state as close as possible before being affected by noise.
To do so, the proposed control procedure consists of two parts: before and after the noise channel parts. Before the noise channel, we apply the pre-weak measurement to gain some information about the initial state. Then the feed-forward operations are used to change the state in a way to reduce the effects of the noise. Amplitude damping leaves the ground states unchanged and decays the excited states by its decaying rate. Hence, the feed-forward operations need to bring the states close to ground state of the noise channel. After the noise channel, the reversed operations are applied to retrieve the information of the initial state.
The schematic diagram of the feed-forward recovery control is given in Fig. \ref{fig1}, which consists of five steps. The details are explained as follows.


Step 1: In the first step we need to obtain some information about the initial state by performing pre-weak measurement as: $\Pi _{00} =M_{00}^{\dag } M_{00} $, $\Pi _{01} =M_{01}^{\dag } M_{01} $, $\Pi _{10} =M_{10}^{\dag } M_{10} $ and $\Pi _{11} =M_{11}^{\dag } M_{11} $ where the measurement operators $M_{ij} (i,j=0,1)$ are given in Table \ref{tab1}, and $p \in \left[ {0,1} \right]$ is the pre-weak measurement strength.

Step 2: The feed-forward operation is applied based on the result of the pre-weak measurement. The feed-forward operation brings the qubit closer to the ground state and makes them less vulnerable to the amplitude damping. The feed-forward operators are given in Table. \ref{tab1}. When the result according to $M_{00} $ is acquired, the system is in the ground state and it is immune to the amplitude damping. We apply the identity operator as $F_{00} $ to keep the state unchanged. As the result according to $M_{01} $ is acquired, $F_{01} $ is chosen as the feed-forward operation. To better understand the effects of feed-forward operation $F_{01} $, we assume the state ${\left| \psi  \right\rangle} =\alpha {\left| 00 \right\rangle} +\beta {\left| 01 \right\rangle} +\gamma {\left| 10 \right\rangle} +\delta {\left| 11 \right\rangle} $. By applying $F_{01} $, it changes to ${\left| \psi ' \right\rangle} =\beta {\left| 00 \right\rangle} +\alpha {\left| 01 \right\rangle} +\delta {\left| 10 \right\rangle} +\gamma {\left| 11 \right\rangle} $.  In such a scenario, by applying $F_{01} $ and replacing the $\beta $ and $\alpha $, we make the state closer to ground state. 

Step 3: The Two-qubit goes through the amplitude damping noise channel. The amplitude damping of a single qubit can be represented by Kraus operators as $e_{0} =\left(\begin{array}{cc} {1} & {0} \\ {0} & {\sqrt{1-r} } \end{array}\right)$, $e_{1} =\left(\begin{array}{cc} {0} & {\sqrt{r} } \\ {0} & {0} \end{array}\right)$, where $r\in \left[0,1\right]$ is the possibility of decaying of the excited state \cite{lab17}. For a two-qubit we assume that amplitude damping occurs for both qubits locally and independently but with the same decaying rate $r=r_{1} =r_{2} $. Therefore the amplitude-damping process for the whole two-qubit system can be described by four Kraus operators $\left(e_{m\, n} ,\, \, m,n=0,1\right)$ as:
\begin{equation} \label{Eq1}
\begin{array}{l} {e_{00} =e_{0} \otimes e_{0} =\left(\begin{array}{cccc} {1} & {0} & {0} & {0} \\ {0} & {\sqrt{1-r} } & {0} & {0} \\ {0} & {0} & {\sqrt{1-r} } & {0} \\ {0} & {0} & {0} & {1-r} \end{array}\right)} \\ {e_{01} =e_{0} \otimes e_{1} =\left(\begin{array}{cccc} {0} & {\sqrt{r} } & {0} & {0} \\ {0} & {0} & {0} & {0} \\ {0} & {0} & {0} & {\sqrt{r} \sqrt{1-r} } \\ {0} & {0} & {0} & {0} \end{array}\right)} \\ {\, e_{10} =e_{1} \otimes e_{0} =\left(\begin{array}{cccc} {0} & {0} & {\sqrt{r} } & {0} \\ {0} & {0} & {0} & {\sqrt{r} \sqrt{1-r} } \\ {0} & {0} & {0} & {0} \\ {0} & {0} & {0} & {0} \end{array}\right)}\\ {e_{11} =e_{1} \otimes e_{1} =\left(\begin{array}{cccc} {0} & {0} & {0} & {r} \\ {0} & {0} & {0} & {0} \\ {0} & {0} & {0} & {0} \\ {0} & {0} & {0} & {0} \end{array}\right)} \end{array}
\end{equation}

Step 4: After the noise channel, the reversed operations $F_{ij} $, which are the same as the ones in step 2, are applied based on the feed-forward operations.

Step 5: We retrieve the information of the initial state by means of post-weak measurement in {a} way that $M_{ij} N_{ij} $ is almost proportionate to $I$. The post-weak measurement is an incomplete measurement with measurement operators given in Table \ref{tab1}, and $q \in \left[ {0,1} \right]$  is the post-weak measurement strength.

The weak measurement operators and corresponding feed-forward operations are given in Table \ref{tab1}.

\begin{table*}[ht]

\caption{Measurement operators and control operations used in two-qubit feed-forward recovery control.}\label{tab1}
\centering
\begin{tabular}{|c|c|c|}
\hline
Pre-weak measurement & Unitary operation & Post-weak measurement  \\
\hline

$M_{00} =\left(\begin{array}{cccc} {p} & {0} & {0} & {0} \\ {0} & {\sqrt{p} \sqrt{1-p} } & {0} & {0} \\ {0} & {0} & {\sqrt{p} \sqrt{1-p} } & {0} \\ {0} & {0} & {0} & {1-p} \end{array}\right)$ & $F_{00} =\left(\begin{array}{cccc} {1} & {0} & {0} & {0} \\ {0} & {1} & {0} & {0} \\ {0} & {0} & {1} & {0} \\ {0} & {0} & {0} & {1} \end{array}\right)$ & $N_{00} =\left(\begin{array}{cccc} {1-q} & {0} & {0} & {0} \\ {0} & {\sqrt{1-q} } & {0} & {0} \\ {0} & {0} & {\sqrt{1-q} } & {0} \\ {0} & {0} & {0} & {1} \end{array}\right)$  \\
\hline
$M_{01} =\left(\begin{array}{cccc} {\sqrt{p} \sqrt{1-p} } & {0} & {0} & {0} \\ {0} & {p} & {0} & {0} \\ {0} & {0} & {1-p} & {0} \\ {0} & {0} & {0} & {\sqrt{p} \sqrt{1-p} } \end{array}\right)$ & $F_{01} =\left(\begin{array}{cccc} {0} & {1} & {0} & {0} \\ {1} & {0} & {0} & {0} \\ {0} & {0} & {0} & {1} \\ {0} & {0} & {1} & {0} \end{array}\right)$ & $N_{01} =\left(\begin{array}{cccc} {\sqrt{1-q} } & {0} & {0} & {0} \\ {0} & {1-q} & {0} & {0} \\ {0} & {0} & {1} & {0} \\ {0} & {0} & {0} & {\sqrt{1-q} } \end{array}\right)$ \\
\hline
$M_{10} =\left(\begin{array}{cccc} {\sqrt{p} \sqrt{1-p} } & {0} & {0} & {0} \\ {0} & {1-p} & {0} & {0} \\ {0} & {0} & {p} & {0} \\ {0} & {0} & {0} & {\sqrt{p} \sqrt{1-p} } \end{array}\right)$ & $F_{10} =\left(\begin{array}{cccc} {0} & {0} & {1} & {0} \\ {0} & {0} & {0} & {1} \\ {1} & {0} & {0} & {0} \\ {0} & {1} & {0} & {0} \end{array}\right)$ & $N_{10} =\left(\begin{array}{cccc} {\sqrt{1-q} } & {0} & {0} & {0} \\ {0} & {1} & {0} & {0} \\ {0} & {0} & {1-q} & {0} \\ {0} & {0} & {0} & {\sqrt{1-q} } \end{array}\right)$ \\ \hline
$M_{11} =\left(\begin{array}{cccc} {1-p} & {0} & {0} & {0} \\ {0} & {\sqrt{p} \sqrt{1-p} } & {0} & {0} \\ {0} & {0} & {\sqrt{p} \sqrt{1-p} } & {0} \\ {0} & {0} & {0} & {p} \end{array}\right)$\newline  & $F_{11} =\left(\begin{array}{cccc} {0} & {0} & {0} & {1} \\ {0} & {0} & {1} & {0} \\ {0} & {1} & {0} & {0} \\ {1} & {0} & {0} & {0} \end{array}\right)$ & $N_{11} =\left(\begin{array}{cccc} {1} & {0} & {0} & {0} \\ {0} & {\sqrt{1-q} } & {0} & {0} \\ {0} & {0} & {\sqrt{1-q} } & {0} \\ {0} & {0} & {0} & {1-q} \end{array}\right)$ \\ \hline

\end{tabular}

\end{table*}

\section{III. Feed-forward recovery control for two-qubit pure initial state} \label{sec3}

Suppose that the quantum system consists of two-qubits, which is in an arbitrary pure initial state as:
\begin{equation}\label{eq2}
 {\left| \psi _{in}  \right\rangle} =\alpha {\left| 00 \right\rangle} +\beta {\left| 01 \right\rangle} +\gamma {\left| 10 \right\rangle} +\delta {\left| 11 \right\rangle}
\end{equation}

We want to protect this state from the noise by means of the feed-forward recovery control based on the following five steps.

Step 1: In the first step the pre-weak measurement is applied. The state of the system ${\left| \psi _{in}  \right\rangle} $ after being measured by the pre-weak measurements $M_{ij} $ in Table 1 becomes ${\left| \psi _{M_{ij} }  \right\rangle} $ as
\begin{equation} \label{eq3}
{\left| \psi _{M_{ij} }  \right\rangle} =\frac{M_{ij} {\left| \psi _{in}  \right\rangle} }{\sqrt{{\left\langle \psi _{in}  \right|} \left. M_{ij}^{\dag } M_{ij} \right|\left. \psi _{in} \right\rangle } }
\end{equation}
with probability $g_{M_{ij} } ={\left\langle \psi _{in}  \right|} \left. M_{ij}^{\dag } M_{ij} \right|\left. \psi _{in} \right\rangle $.

 Step 2: The feed-forward operation $F_{ij} $ in Table \ref{tab1} is applied based on the result of the pre-weak measurement. The state of the system ${\left| \psi _{M_{ij} }  \right\rangle} $ after the feed-forward operation is given by ${\left| \psi _{F_{ij} }  \right\rangle} $:
\begin{equation} \label{eq4)}
{\left| \psi _{F_{ij} }  \right\rangle} =F_{ij} {\left| \psi _{M_{ij} }  \right\rangle}
\end{equation}

Step 3: The Two-qubit goes through the noisy channel. The state of the system ${\left| \psi _{F_{ij} }  \right\rangle} $ after passing through the noise channel is not pure anymore. But to make the calculation more manageable, we use a mathematical technique called unravelling. So the qubit trajectories can be divided into two parts, `jump' and `no jump' trajectories, and each qubit can jump to state ${\left| 0 \right\rangle} $, or `no jump' happens. If both qubits jump, the system state becomes ${\left| \psi ^{e_{11} }  \right\rangle} ={\left| 00 \right\rangle} $ with probability $g_{e_{11} } ={\left\langle \psi _{F_{ij} }  \right|} \left. e_{11} ^{\dag } e_{11} \right|\left. \psi _{F_{ij} } \right\rangle $, which is non-invertible. If `no jump' scenario happens for at least one of the qubits, the state of the system transforms to ${\left| \psi ^{e_{mn} }  \right\rangle} =\frac{e_{mn} {\left| \psi _{F_{ij} }  \right\rangle} }{\sqrt{{\left\langle \psi _{F_{ij} }  \right|} \left. e_{mn} ^{\dag } e_{mn} \right|\left. \psi _{F_{ij} } \right\rangle } } $ with probability $g_{e_{mn} } ={\left\langle \psi _{F_{ij} }  \right|} \left. e_{mn} ^{\dag } e_{mn} \right|\left. \psi _{F_{ij} } \right\rangle $ and we can recover the state of the system.

Step 4: After the noise channel, the reversed operations $F_{ij} $ based on the feed-forward operations applied. The state of the system after using the reversed operation is ${\left| \psi _{F_{ij} }^{e_{mn} }  \right\rangle} $ :
\begin{equation} \label{eq5)}
{\left| \psi _{F_{ij} }^{e_{mn} }  \right\rangle} =F_{ij} {\left| \psi ^{e_{mn} }  \right\rangle}
\end{equation}

Step 5: At last, we measure the state of the system by post-weak measurement operators $N_{ij} $. The state of the system after being measured by the post-weak measurement is presented as:
\begin{equation} \label{eq6)}
{\left| \psi _{N_{ij} }^{e_{m\, n} }  \right\rangle} =\frac{N_{ij} {\left| \psi _{F_{ij} }^{e_{m\, n} }  \right\rangle} }{\sqrt{{\left\langle \psi _{F_{ij} }^{e_{m\, n} }  \right|} \left. N_{ij}^{\dag } N_{ij} \right|\left. \psi _{F_{ij} }^{e_{m\, n} } \right\rangle } }
\end{equation}
with probability $g_{N_{ij} }^{e_{m\, n} } ={\left\langle \psi _{F_{ij} }^{e_{m\, n} }  \right|} \left. N_{ij}^{\dag } N_{ij} \right|\left. \psi _{F_{ij} }^{e_{m\, n} } \right\rangle $.

Since we consider the damped state in two scenarios, `jump' and `no jump', the final state of the system corresponding to $M_{ij} $ is given by:
\begin{equation} \label{eq7)}
\rho _{M_{ij} }^{fin} =\sum _{m,n=0}^{1}\frac{g_{N_{ij} }^{e_{m\, n} } {\left| \psi _{N_{ij} }^{e_{m\, n} }  \right\rangle} {\left\langle \psi _{N_{ij} }^{e_{m\, n} }  \right|} }{\sqrt{g_{N_{ij} }^{e_{m\, ns} } } }
\end{equation}
with the success probability $g_{M_{ij} }^{fin} $:
\begin{equation} \label{eq8)}
g_{M_{ij} }^{fin} =\sum _{m,n=0}^{1}g_{N_{ij} }^{e_{m\, n} }
\end{equation}

To calculate the performance of the recovery control process, we use the fidelity $Fid_{M_{ij} }$ between the initial state ${\left| \psi _{in}  \right\rangle} $ and the final state $\rho _{M_{ij} }^{fin} $ that corresponds to each weak measurement operator as: $Fid_{M_{ij} } ={\left\langle \psi _{in}  \right|} \left. \rho _{M_{ij} }^{fin} \right|\left. \psi _{in} \right\rangle $. The final total fidelity $Fid_{total} $ and final total probability $g_{total}^{} $ after the whole process of recovery control, respectively, are: 
\begin{equation} \label{eq9}
\resizebox{.5 \textwidth}{!}
{$Fid_{total} =\frac{g_{M_{00} }^{fin} Fid_{M_{00} } +g_{M_{01} }^{fin} Fid_{M_{01} } +g_{M_{10} }^{fin} Fid_{M_{10} } +g_{M_{11} }^{fin} Fid_{M_{11} } }{g_{M_{00} }^{fin} +g_{M_{01} }^{fin} +g_{M_{10} }^{fin} +g_{M_{11} }^{fin} } $ }
\end{equation}
\[g_{total}^{} =g_{M_{00} }^{fin} +g_{M_{01} }^{fin} +g_{M_{10} }^{fin} +g_{M_{11} }^{fin} \]

To better understand the control process, we present an analytic expression only for measurement operator $M_{00} $. All the states corresponding to all measurement operators can be analytically obtained, for brevity we do not bring them in this paper although.

In the first step, we apply a pre-weak measurement defined by $M_{ij} $. Let us consider the result corresponding to  $M_{00} $ is acquired. The state of the system ${\left| \psi _{in}  \right\rangle} $ in Eq. \eqref{eq2} after being measured by $M_{00} $ becomes
\begin{equation} \label{eq10)}
\begin{aligned}
{\left| \psi _{M_{00} }  \right\rangle} =M_{00} {\left| \psi _{in}  \right\rangle}&=\frac{1}{\sqrt{g_{M_{00} } } } \bigr(\alpha p{\left| 00 \right\rangle} +\beta \sqrt{p} \sqrt{1-p} {\left| 01 \right\rangle}  \\ &+\gamma \sqrt{p} \sqrt{1-p} {\left| 10 \right\rangle} +\delta \sqrt{1-p} {\left| 11 \right\rangle} \bigl)
\end{aligned}
\end{equation}
where $g_{M_{00} } =\alpha ^{2} p^{2} +\beta ^{2} p(1-p)+\gamma ^{2} p(1-p)+\delta ^{2} \left(p-1\right)^{2} $ is the probability of achieving the result according to measurement operator $M_{00} $.

Since the result that corresponds to $M_{00} $ happens, before noise we choose the feed-forward operation $F_{00} $, and do-nothing. ${\left| \psi _{F_{00} }  \right\rangle} $ is the state of the system after applying the feed-forward operation:
\begin{equation} \label{eq11)}
{\left| \psi _{F_{00} }  \right\rangle} =F_{00} {\left| \psi _{M_{00} }  \right\rangle} ={\left| \psi _{M_{00} }  \right\rangle}
\end{equation}

Now the two-qubit enters the noise channel. As we explained before the state of the system `jumps' into ${\left| 00 \right\rangle} $ state or `no jump' happens.  The `jump' scenario for both qubits happens with probability $g_{e_{11} } ={\left\langle \psi _{F_{ij} }  \right|} \left. e_{11} ^{\dag } e_{11} \right|\left. \psi _{F_{ij} } \right\rangle =\left|\delta \right|^{2} r^{2} (p-1)^{2} $, and the state of the system becomes ${\left| \psi ^{e_{11} }  \right\rangle} ={\left| 00 \right\rangle} $. Also, `no jump' scenario for both qubits transfers the state into ${\left| \psi ^{e_{00} }  \right\rangle} $ as:
\begin{equation} \label{eq12)}
\begin{aligned}
{\left| \psi _{M_{01} }^{e_{00} }  \right\rangle} &=\frac{1}{\sqrt{g_{e_{00} } } } \bigl(\alpha p{\left| 00 \right\rangle} \\&+\beta \sqrt{p} \sqrt{1-p} \sqrt{1-r} {\left| 01 \right\rangle}\\ &+ \gamma \sqrt{p} \sqrt{1-p} \sqrt{1-r} {\left| 10 \right\rangle} \\ &+\delta (1-p)(1-r){\left| 11 \right\rangle} \bigr)
\end{aligned}
\end{equation}
where $g_{e_{00} } =\left|\alpha \right|^{2} p^{2} +(\left|\beta \right|^{2} +\left|\gamma \right|^{2} )\left(p(1-p)\left(1-r\right)\right)+\left|\delta \right|^{2} (1-p)^{2} \left(1-r\right)^{2} $ is the probability of no jumping of both qubits after the noise channel. The state of the system in the case that just one of the qubits jumps is analysed in Appendix A. Here we calculate the states which both qubits jump or no jump.
After the noise channel, we make the reversed operation $F_{00} $. Hence the state of the system from `no jumping' trajectory becomes ${\left| \psi _{F_{00} }^{e_{00} }  \right\rangle} =F_{00} {\left| \psi ^{e_{00} }  \right\rangle} ={\left| \psi ^{e_{00} }  \right\rangle} $, and the `jump' trajectory becomes ${\left| \psi _{F_{00} }^{e_{11} }  \right\rangle} =F_{00} {\left| \psi ^{e_{11} }  \right\rangle} ={\left| 00 \right\rangle} $.

At last the post-weak measurement is applied by using the measurement operator $N_{00}$. The state of the system from `no jumping' scenario becomes:
\begin{equation}\label{eq13)}
\begin{aligned}
{\left| \psi _{N_{00} }^{e_{00} }  \right\rangle} &=\frac{1}{\sqrt{g_{N_{00} } } }  \bigl(\alpha p\left(1-q\right){\left| 00 \right\rangle} \\&+  \beta  \sqrt{p} \sqrt{1-p} \sqrt{1-r} \sqrt{1-q} {\left| 01 \right\rangle}\\ &+\gamma \sqrt{p} \sqrt{1-p} \sqrt{1-r} \sqrt{1-q} {\left| 10 \right\rangle} \\&+\delta (1-p)(1-r){\left| 11 \right\rangle} \bigr)
\end{aligned}
\end{equation}

\noindent with probability $g_{N_{00} }^{e_{00} } =\left|\alpha \right|^{2} p^{2} \left(1-q\right)^{2} +(\left|\beta \right|^{2} +\left|\gamma \right|^{2} )\left(p(1-p)\left(1-r\right)\left(1-q\right)\right)+\left|\delta \right|^{2} (1-p)^{2} \left(1-r\right)^{2} $.

The jumping state after post-weak measurement becomes ${\left| \psi _{N_{00} }^{e_{11} }  \right\rangle} ={\left| 00 \right\rangle} $ with probability $g_{N_{00} }^{e_{11} } =\left|\delta \right|^{2} r^{2} (1-p)^{2} (1-q)^{2} $. The final state of the system corresponds to measurement operator $M_{00} $ after passing through the whole control procedure is given by:

\begin{equation} \label{eq14)}
\begin{aligned}
&\rho _{M_{00} }^{fin} =\frac{1}{g_{N_{00} }^{e_{00} } +g_{N_{00} }^{e_{01} } +g_{N_{00} }^{e_{10} } +g_{N_{00} }^{e_{11} } } \Bigr(g_{N_{00} }^{e_{00} } {\left| \psi _{N_{00} }^{e_{00} }  \right\rangle} {\left\langle \psi _{N_{00} }^{e_{00} }  \right|} \\&+g_{N_{00} }^{e_{01} } {\left| \psi _{N_{00} }^{e_{01} }  \right\rangle} {\left\langle \psi _{N_{00} }^{e_{01} }  \right|} +g_{N_{00} }^{e_{10} } {\left| \psi _{N_{00} }^{e_{10} }  \right\rangle} {\left\langle \psi _{N_{00} }^{e_{10} }  \right|} +g_{N_{00} }^{e_{11} } {\left| 00 \right\rangle} {\left\langle 00 \right|}\Bigl)
\end{aligned}
\end{equation}
where  ${\left| \psi _{N_{00} }^{e_{01} }  \right\rangle} {\left\langle \psi _{N_{00} }^{e_{01} }  \right|} ,\, {\left| \psi _{N_{00} }^{e_{10} }  \right\rangle} {\left\langle \psi _{N_{00} }^{e_{10} }  \right|} ,\, g_{N_{00} }^{e_{01} } $, and $\, g_{N_{00} }^{e_{10} } $ are presented in the appendix A.

\section{IV. Feed-forward recovery control for two-qubit mixed initial state} \label{sec4}

The arbitrary two-qubit mixed initial state can be represented in matrix form as:
\begin{equation} \label{eq15)}
\bar{\rho }_{in} =\left(\begin{array}{cccc} {a} & {e} & {f} & {g} \\ {e'} & {b} & {h} & {i} \\ {f'} & {h'} & {c} & {j} \\ {g'} & {i'} & {j'} & {d} \end{array}\right)
\end{equation}

To apply the proposed feed-forward recovery control in the case of the mixed initial state, we use a quantum operation, a completely-positive trace-preserving (CPTP) map, that acts on a two-qubit density matrix. Hence the non-normalized final recovered state corresponds to measurement operator $M_{ij} (i,j=0,1)$ is given as:
\begin{equation} \label{eq16}
\begin{aligned}
C(\bar{\rho }_{M_{ij} }^{fin}) &=\sum _{m,\, n=0}^{1}N_{ij} F_{ij} e_{m\, n} F_{ij} M_{ij} \rho _{in} \left(N_{ij} F_{ij} e_{m\, n} F_{ij} M_{ij} \right)^{\dag }  \\&=\sum _{m,\, n=0}^{1}N_{ij} F_{ij} e_{m\, n} F_{ij} M_{ij} \rho _{in} M_{ij}^{\dag } F_{ij}^{\dag } e_{mn}^{\dag }  F_{ij}^{\dag } N_{ij}^{\dag }
\end{aligned}
\end{equation}
where $N_{ij} $ are the post-weak measurement operators, $F_{ij} $ are the feed-forward operations, $M_{ij} $ are the pre-weak measurement operators given in Table. 1. Moreover, $e_{m\, n} ,m,n=0,1$ are four different Kraus operators for amplitude damping noise given in Eq. \eqref{Eq1}.
The probability for gaining the result $\bar{\rho }_{M_{ij} } $ is the normalization factor of Eq. \eqref{eq16} as:
\begin{equation} \label{eq17}
\bar{g}_{M_{ij} }^{fin} =\sum _{m,\, n=0}^{1}trace\resizebox{.32 \textwidth}{!} {$\left(N_{ij} F_{ij} e_{m\, n} F_{ij} M_{ij} \rho _{in} M_{ij}^{\dag } F_{ij}^{\dag } e_{mn}^{\dag } F_{ij}^{\dag } N_{ij}^{\dag } \right)$}
\end{equation}

 Since the initial and final states are both mixed, we find the final fidelity corresponds to each measurement operator $M_{ij} $ as:
\begin{equation} \label{eq18}
\overline{Fid}_{M_{ij} } =\left[Tr\left(\sqrt{\sqrt{\bar{\rho }_{M_{ij} }^{fin} } \bar{\rho }_{in} \sqrt{\bar{\rho }_{M_{ij} }^{fin} } } \right)\right]^{2}
\end{equation}

The final total fidelity and final total success probability can be defined as Eq. \eqref{eq9}.

\section{V. Complete recovery condition of feed-forward control} \label{sec5}

As we explained in Sec. II, the state of the system after passing through the amplitude damping noise channel will be in `jump' scenario or `no jump' scenario. If one of the qubits or both of them jump, we are not able to retrieve the information of the initial state.  However, when `no jump' scenario happens for both qubits by choosing the appropriate measurement strength for the post-weak measurements, one can make the state of the system after post-weak measurement ${\left| \psi _{N_{ij} }^{e_{00} }  \right\rangle} $, completely same as the initial state ${\left| \psi _{in}  \right\rangle} $.  ${\left| \psi _{N_{ij} }^{e_{00} }  \right\rangle} $ given in Eq. \eqref{eq13)} is the final state of the system for measurement operator $M_{00} $ when no jump happens for both qubits. It should be equal to the initial state of the system given in Eq. \eqref{eq2}. Hence, all the vectors should be equal as: $p\left(1-q\right)=\sqrt{p} \sqrt{1-p} \sqrt{1-r} \sqrt{1-q} =\sqrt{p} \sqrt{1-p} \sqrt{1-r} \sqrt{1-q} =(1-p)(1-r)$

Therefore, the complete recovery condition is:
\begin{equation} \label{eq19}
q=1-\frac{(1-p)(1-r)}{p}
\end{equation}
which is a function of damping probability $r$ and pre-weak measurement strength $p$.
By substituting the post-weak measurement strength $q$ as Eq. \eqref{eq19}, the state of the system given in Eq. \eqref{eq13)} becomes:
\begin{equation} \label{eq20)}
\begin{aligned}
{\left| \widetilde{\psi }_{N_{00} }^{e_{00} }  \right\rangle} &=\frac{1}{\sqrt{\widetilde{g}_{N_{00} }^{e_{00} } } } \Bigr(\alpha (1-p)(1-r){\left| 00 \right\rangle}\\ &+\beta (1-p)(1-r){\left| 01 \right\rangle}\\& +\gamma (1-p)(1-r){\left| 10 \right\rangle} \\&+\delta (1-p)(1-r){\left| 11 \right\rangle} \Bigl) \\&=\frac{(1-p)(1-r)}{\sqrt{\widetilde{g}_{N_{00} }^{e_{00} } } } \left(\alpha {\left| 00 \right\rangle} +\beta {\left| 01 \right\rangle} +\gamma {\left| 10 \right\rangle} +\delta {\left| 11 \right\rangle} \right)
\end{aligned}
\end{equation}
with probability
\begin{equation} \label{eq21)}
\begin{aligned} \widetilde{g}_{N_{00} }^{e_{00} } &=\left|\alpha \right|^{2} (1-p)^{2} \left(1-r\right)^{2} +\left|\beta \right|^{2} (1-p)^{2} \left(1-r\right)^{2} \\&+\left|\gamma \right|^{2} (1-p)^{2} \left(1-r\right)^{2} +\left|\delta \right|^{2} (1-p)^{2} \left(1-r\right)^{2}  \\  &=(1-p)^{2} \left(1-r\right)^{2} \left(\left|\alpha \right|^{2} +\left|\beta \right|^{2} +\left|\gamma \right|^{2} +\left|\delta \right|^{2} \right)\\&=(1-p)^{2} \left(1-r\right)^{2}
\end{aligned}
\end{equation}

Hence, one can see that the final state becomes exactly the same as the initial state:
\begin{equation} \label{eq22)}
\begin{aligned}
{\left| \widetilde{\psi }_{N_{00} }^{e_{00} }  \right\rangle} &=\frac{(1-p)(1-r)}{\sqrt{(1-p)^{2} \left(1-r\right)^{2} } } \resizebox{.23 \textwidth}{!}{$\left(\alpha {\left| 00 \right\rangle} +\beta {\left| 01 \right\rangle} +\gamma {\left| 10 \right\rangle} +\delta {\left| 11 \right\rangle} \right)$}\\&=\alpha {\left| 00 \right\rangle} +\beta {\left| 01 \right\rangle} +\gamma {\left| 10 \right\rangle} +\delta {\left| 11 \right\rangle} ={\left| \psi _{in}  \right\rangle}
\end{aligned}
\end{equation}

By substituting Eq. \eqref{eq19} in Eq. \eqref{eq9} (success probability for pure initial state) or Eq. \eqref{eq17} (success probability for mixed initial state), the complete recovery success probability becomes:
\begin{equation} \label{eq23}
g_{fin}^{total} =\frac{(1-p)^{2} (1-r)^{2} (2p+r-pr)^{2} }{p^{2} }
\end{equation}

\begin{figure}[t]
\centering
\includegraphics[width=7.1cm]{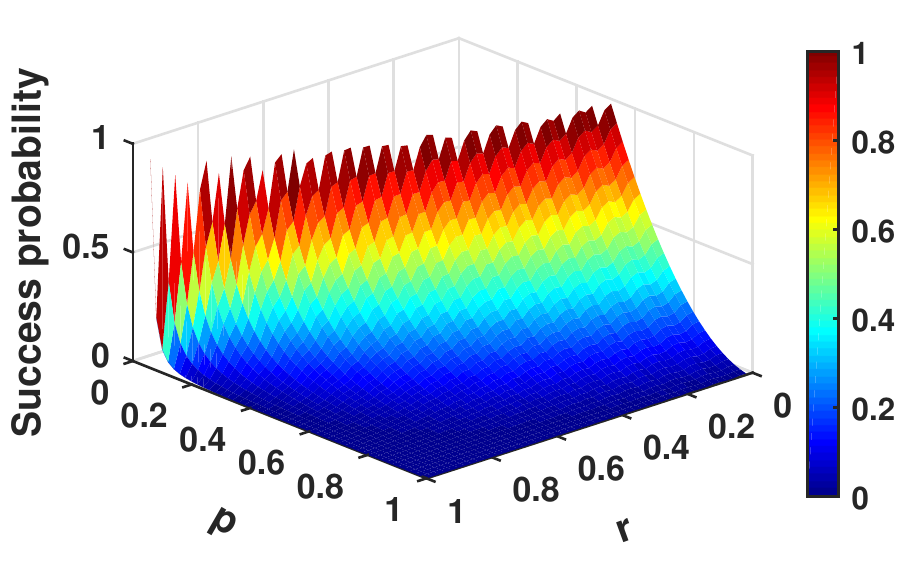}
\caption{Complete recovery total final success probability as a function of pre-weak measurement strength and damping probability.}
\label{fig2}
\end{figure}

According to Eq. \eqref{eq23}, the total final success probability under the complete recovery condition is not a function of the initial state. So the behaviour of the success probability is the same for mixed and pure initial states. In other words, under the complete recovery condition given in Eq. \eqref{eq19} the total final success probability for pure initial states Eq. \eqref{eq9} and mixed initial states Eq. \eqref{eq17} have the same amount as given in Eq. \eqref{eq23}. To study the behaviour of the complete recovery total final success probability, we set the value of post-weak measurement strength $q$ as complete recovery condition in Eq. \eqref{eq19}, and use Monte-Carlo method over a large ensemble of two-qubit initial states. The complete recovery total final success probability as a function of pre-weak measurement strength $p$ and damping probability $r$ for pure and mixed initial states via Monte-Carlo method is given in Fig. \ref{fig2}.

As can be seen from Fig. \ref{fig2}, in complete recovery control the success probability has the most significant value while the pre-weak measurement strength is so weak. By increasing the amount of pre-weak measurement strength the amount of total success probability decreases and tends to zero. We note that in complete recovery control, according to Eq. \eqref{eq19} $q$ is the function of pre-weak measurement $p$ and damping probability $r$. Since $q\ge 0$ the lowest amount of $p$ depends on damping probability $r$. That's why in Fig. \ref{fig2} there are some gaps in the success probability for the small pre-weak measurement strength $p$.

\begin{figure}[!h]
    \centering
    \begin{minipage}[t]{0.49\textwidth}
        \centering{\includegraphics[width=\textwidth]{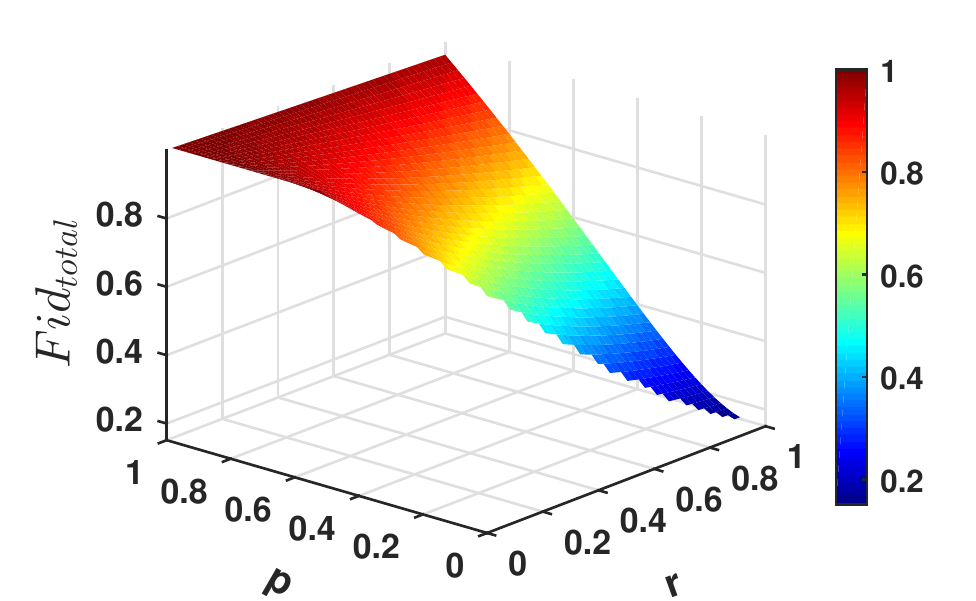}}
        \centering{(a) Pure initial states}\label{fig3a}
    \end{minipage}
    \hfill
    \begin{minipage}[t]{0.49\textwidth}
        \centering{\includegraphics[width=\textwidth]{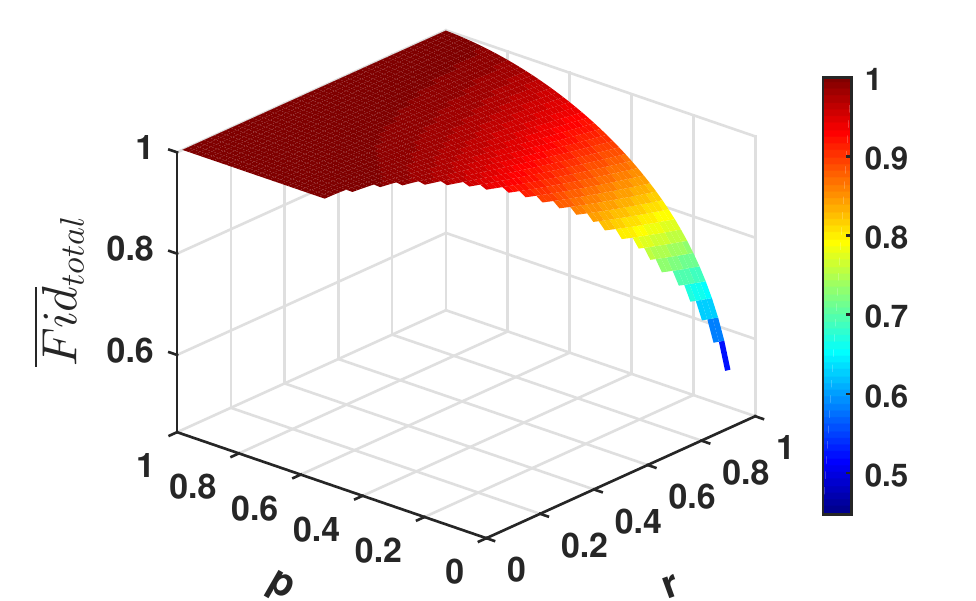}}
        \centering{(b) Mixed initial states initial states}\label{fig3b}
    \end{minipage}
\caption{ Complete recovery total final fidelity as a function of pre-weak measurement  and damping probability  for mixed initial state.}\label{fig3}
\end{figure}

The final expression for complete recovery fidelity is complicated, so the behaviour of the fidelity is explained in numerical simulations. The behaviour of the complete recovery total final success probability as a function of pre-weak measurement strength and damping probability for pure and mixed initial states via Monte-Carlo method are given in Fig. \ref{fig3}. Fig. 3(a) is the complete recovery total final fidelity for pure initial states as in Eq. \eqref{eq9} and Fig. 3(b) is the complete recovery total final fidelity for mixed initial states by substituting Eq. \eqref{eq18} in Eq. \eqref{eq9}.

\begin{figure*}[ht!]
    \centering
    \begin{minipage}[t]{0.49\textwidth}
        \centerline{\includegraphics[width=0.9\textwidth]{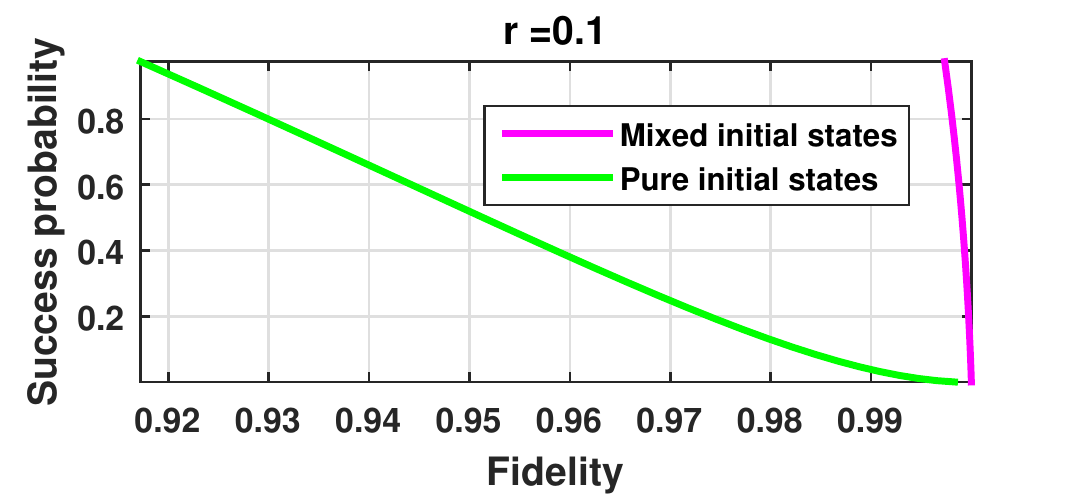}}
        \centerline{(a) r=0.1}\label{fig4a}
    \end{minipage}
    \hfill
    \begin{minipage}[t]{0.49\textwidth}
         \centerline{\includegraphics[width=0.9\textwidth]{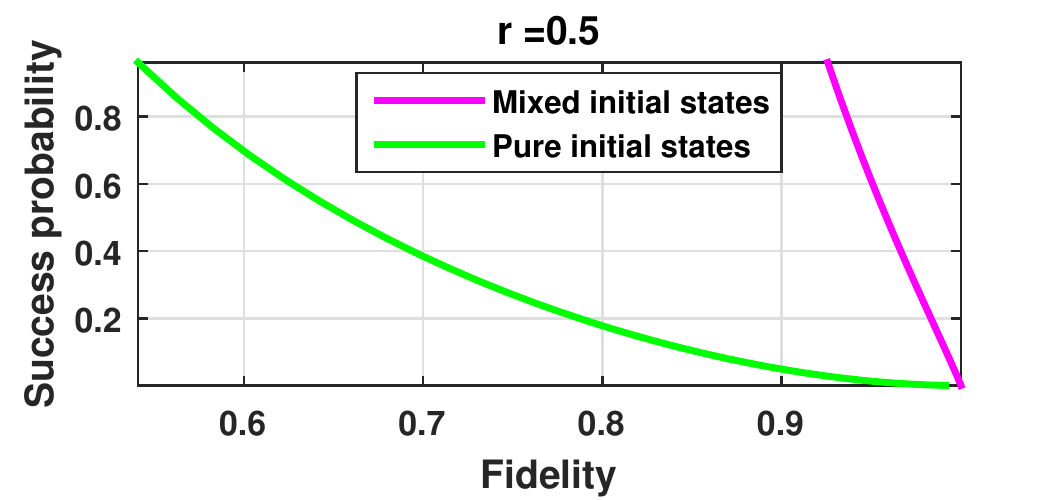}}
          \centerline{(b) r=0.6}\label{fig4b}
    \end{minipage}
    \hfill
    \begin{minipage}[t]{0.49\textwidth}
          \centerline{\includegraphics[width=0.9\textwidth]{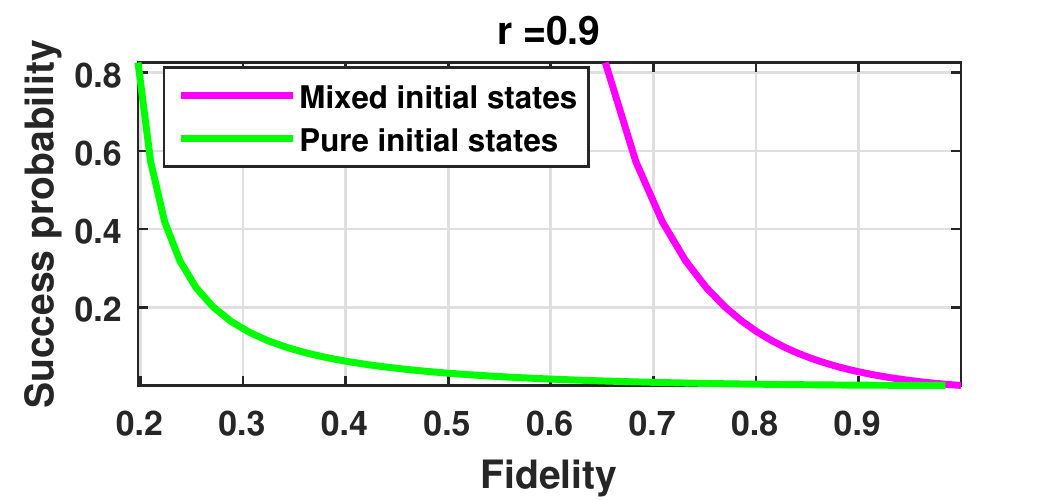}}
          \centerline{(c) r=0.9}\label{fig4c}
    \end{minipage}
\caption{ The relation between total final fidelity and total success probability under complete recovery condition for (r=0. 01, 0.5 and 0.9)  . Green curves correspond to pure initial states and magnet curves to mixed initial states.}\label{fig4}
\end{figure*}
Total complete recovery fidelity increases by increasing the amount of pre-weak measurement strength as shown in Fig. \ref{fig3} for both pure and mixed initial states. However, the fidelity of the complete recovery control is better in case of mixed initial states. It can be seen from Fig. 3(b) that the fidelity of mixed initial states is more than 99\% for all amounts of damping probability by choosing the appropriate amount of pre-weak measurement.

Furthermore, to show the effectiveness of our recovery control for pure and mixed initial state, Fig. \ref{fig4} shows the relation between total final fidelity and total success probability via Monte Carlo method over $10^{4} $ iteration of the two-qubit mixed and pure initial states. Once more, the complete recovery total final fidelity for pure initial states is given in Eq. \eqref{eq9}, and the complete recovery total final fidelity for mixed initial states calculated by substituting Eq. \eqref{eq18} in Eq. \eqref{eq9}. The complete recovery success probability is given in Eq. \eqref{eq19}. In each plot, the damping probability $r$ is fixed as (r=0. 01, 0.5 and 0.9), and the pre-weak measurement strength $p$ changes from 0 to 1.

From Fig \ref{fig4}, one can see that the higher the fidelity is, the lower success probability becomes, and vice versa. However, even for heavy damping probability our control can protect two-qubit pure and mixed states from noise. By comparing the relation between total final success probability and total final fidelity of mixed and pure initial states, one can see the feed-forward operation has the same success probability but better fidelity for mixed initial states.

\begin{figure}[h!]
\centering
\includegraphics[width=7.1cm]{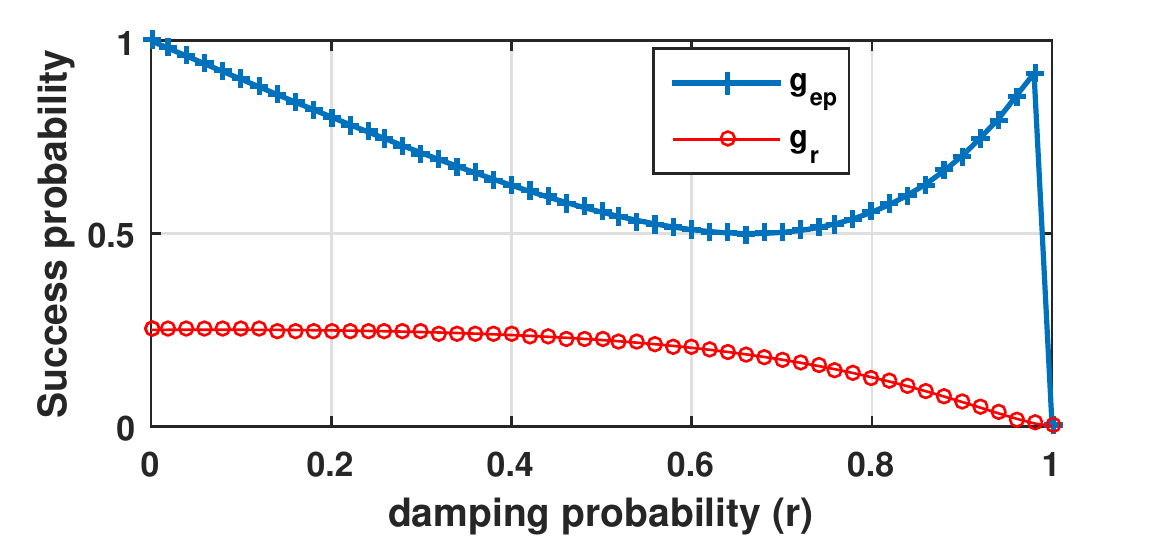}
\caption{Success probability as a function of damping probability $r$ via Monte-Carlo method. The red curve is the success probability for protecting scheme in ref. \cite{lab15}; and blue curve is the fidelity of our scheme with complete recovery condition.}
\label{fig5}
\end{figure}

In order to compare the effectiveness of our recovery control for protecting two-qubit against amplitude damping, we demonstrate the fidelity and success probability in complete recovery condition and the one in ref. \cite{lab15}. We note that, in ref. \cite{lab15} the amplitude damping caused by weak measurement has been considered, which is equivalent to `no jump' scenario for both qubits in our paper. Hence, we only consider the case of `no jump' scenario. For each amount of damping probability $r$ the best pre-weak measurement $p$ is chosen for calculating the total fidelity and success probability of the complete recovery control. As showed in Fig. \ref{fig2} and Fig. \ref{fig3} the highest fidelity is obtained with stronger pre-weak measurement and highest success probability is obtained with weaker pre-weak measurement. We generate arbitrary two-qubit pure states density matrices via extensive Monte-Carlo method over $10^{4} $ iteration. Success probability as a function of damping probability for our complete recovery control given in Eq. \eqref{eq9} and recovery control in ref. \cite{lab15} are shown in Fig. \ref{fig5}.

From Fig \ref{fig5} we can see that by choosing the appropriate amount of pre-weak measurement strength $p$ our control with complete recovery condition makes significant improvement for success probability even for high amount of damping probability.

\begin{figure}[h]
\centering
\includegraphics[width=7.1cm]{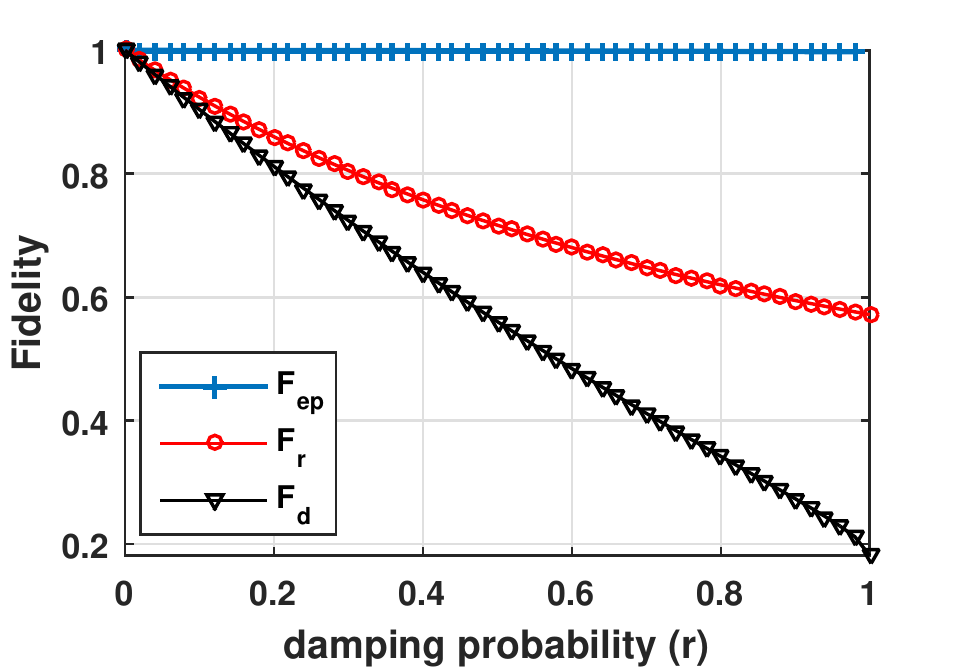}
\caption{Fidelity as a function of damping probability $r$ via Monte-Carlo method. The black curve is the fidelity without any control field, red curve is the fidelity for protecting scheme in ref. \cite{lab15}; and blue curve is the fidelity of our scheme with complete recovery condition.}
\label{fig6}
\end{figure}

In addition, the fidelity as a function of damping probability via Monte-Carlo method is shown in Fig. \ref{fig6}. $F_{d} $ is the fidelity without any control field, fidelity between damped state and the initial state; $F_{r} $ corresponds to the fidelity of the recovery control in ref. \cite{lab15}; and $F_{ep} $ is the fidelity of our control with complete recovery condition. Since we choose the strongest pre-weak measurement strength, fidelity of our control is close to 1 for all amounts of damping probability.

\begin{figure}[h]
\centering
\includegraphics[width=7.1cm]{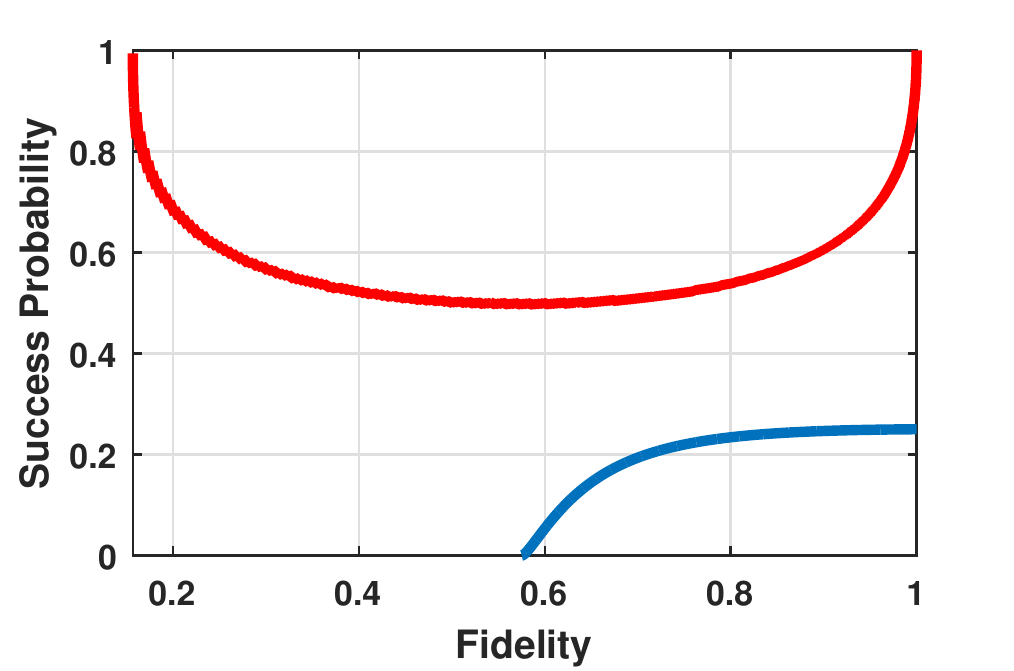}
\caption{The relation between fidelity and success probability via Monte-Carlo method. The red curve corresponds to our scheme and blue curve is for the one in ref. \cite{lab15}.}
\label{fig7}
\end{figure}

 Also, to compare the amount of success probability (fidelity) for the given fidelity (success probability), we plot the relation between fidelity and success probability for our recovery control with complete recovery condition and the recovery control in ref. \cite{lab15} in Fig. \ref{fig7}. To make sure that our comparison is not just for some suitable states we use Monte-Carlo method over $10^{4} $ iteration. The amount of damping probability changes from 0 to 1, and each point is the corresponding amount of fidelity and success probability. The pre-weak measurement strength was chosen to be the smallest amount for each damping probability amount based on Eq. \eqref{eq19}.

As Fig. \ref{fig7} depicted our recovery control always has significant improvement in terms of success probability (fidelity) for the given fidelity (success probability).

\section{VI. Experiments and Discussions} \label{sec6}

In this Section, we consider the optimal recovery control for two-qubit in which the variables of pre-weak measurement strength $p$ and post-weak measurement strength $q\, $ can have dependent amounts.

The final expression for total success probability as a function of pre-weak measurement strength $p$, post-weak measurement strength $q\, $ and damping probability $r$ is:
\begin{equation} \label{eq24}
\bar{g}_{fin}^{total} =\left(pq+qr-pqr-1\right)^{2}
\end{equation}

\begin{figure}[h]
\centering
\includegraphics[width=8cm]{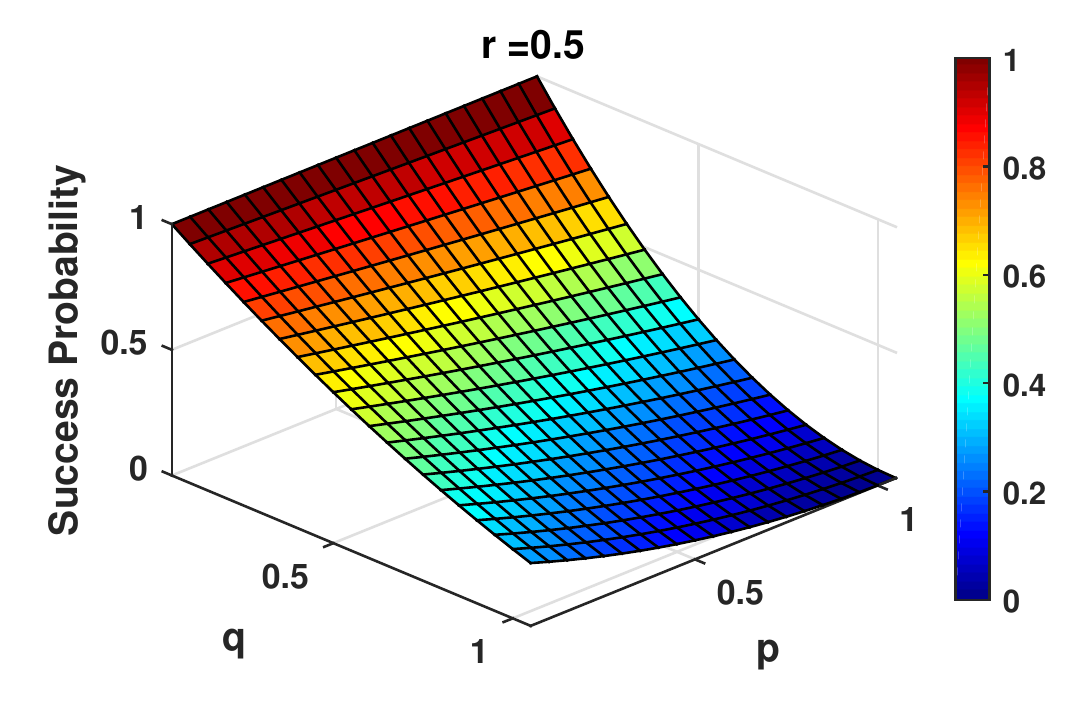}
\caption{General recovery control total success probability as a function of pre-weak measurement strength $p$ and post-weak measurement strength $q$ for fixed amount of  $r=0.5$.}
\label{fig8}
\end{figure}

According to Eq. \eqref{eq24} the total success probability in the general recovery scheme, is not a function of the initial state. Hence, the behaviors of success probability for pure and mixed initial states are the same. Fig. \ref{fig8} depicted the total success probability given in Eq. \eqref{eq24}, as a function of pre-weak measurement strength $p$ and post-weak measurement strength $q\, $ for fixed amount of $r=0.5$.

As Fig. \ref{fig8} depicted, in general the recovery control post-weak measurement strength has most effects on the behavior of success probability. By choosing weaker post-weak measurement strength one can gain a higher success probability.

\begin{figure*}[!ht]
    \centering
    \begin{minipage}[t]{0.3\textwidth}
        \centerline{\includegraphics[width=\textwidth]{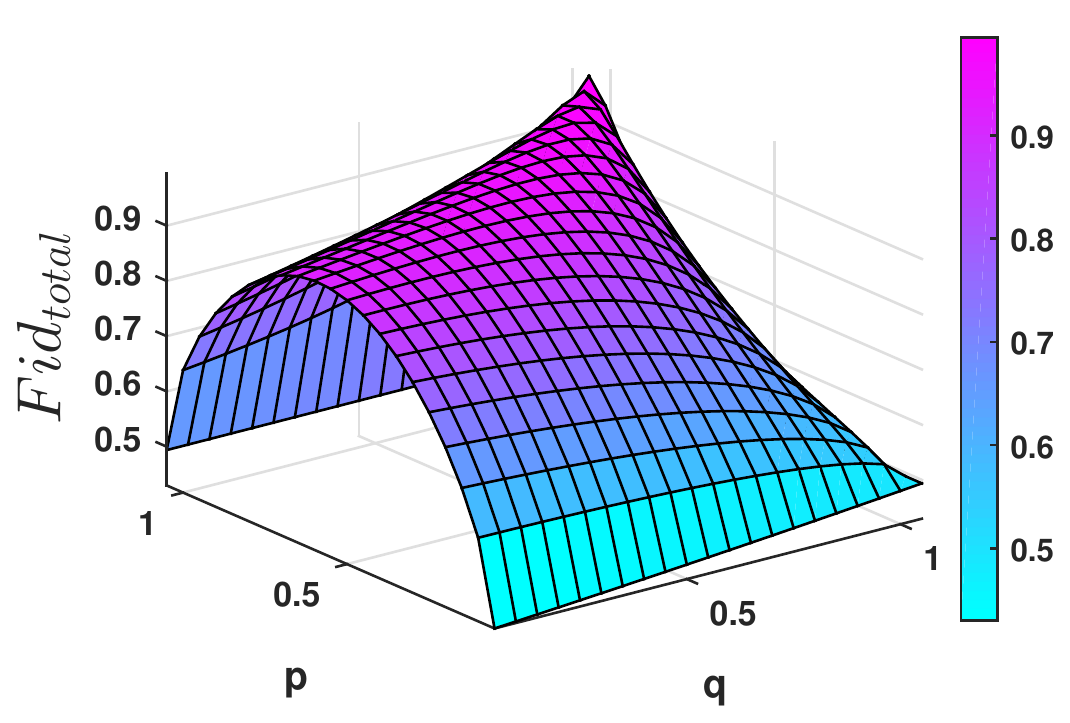}}
        \centerline{(a) Pure initial states ($r=0.1$)}
    \end{minipage}
    \hfill
    \begin{minipage}[t]{0.3\textwidth}
        \centering{\includegraphics[width=\textwidth]{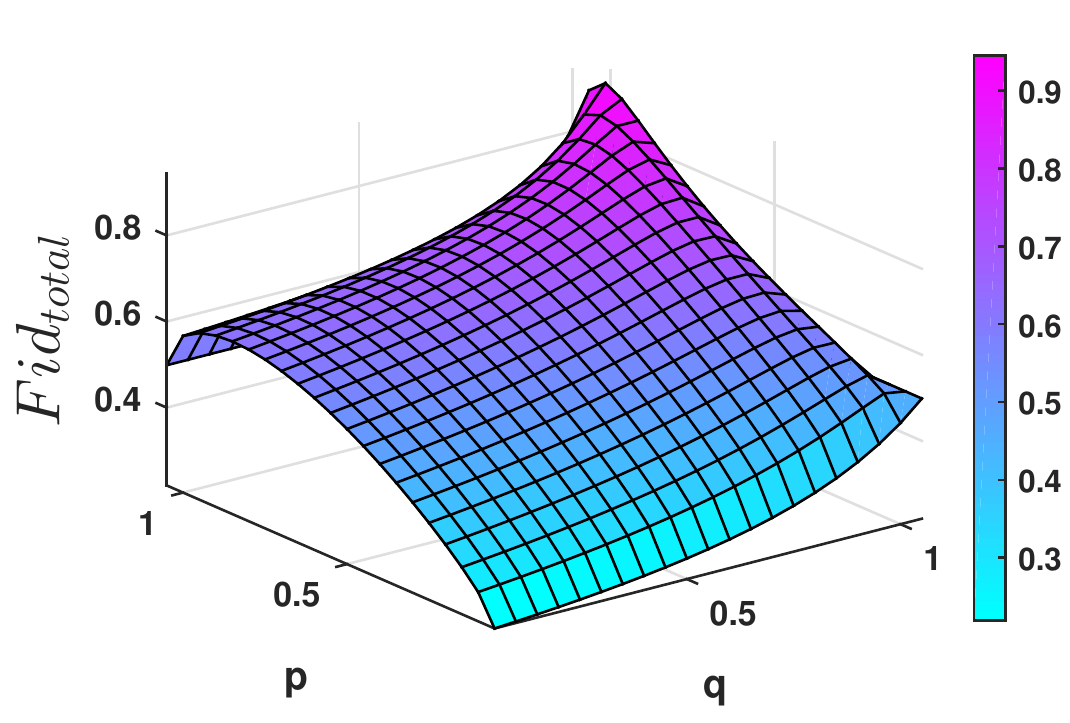}}
        \centering{(b) Pure initial states ($r=0.6$)}
    \end{minipage}
    \hfill
    \begin{minipage}[t]{0.3\textwidth}
        \centering{\includegraphics[width=\textwidth]{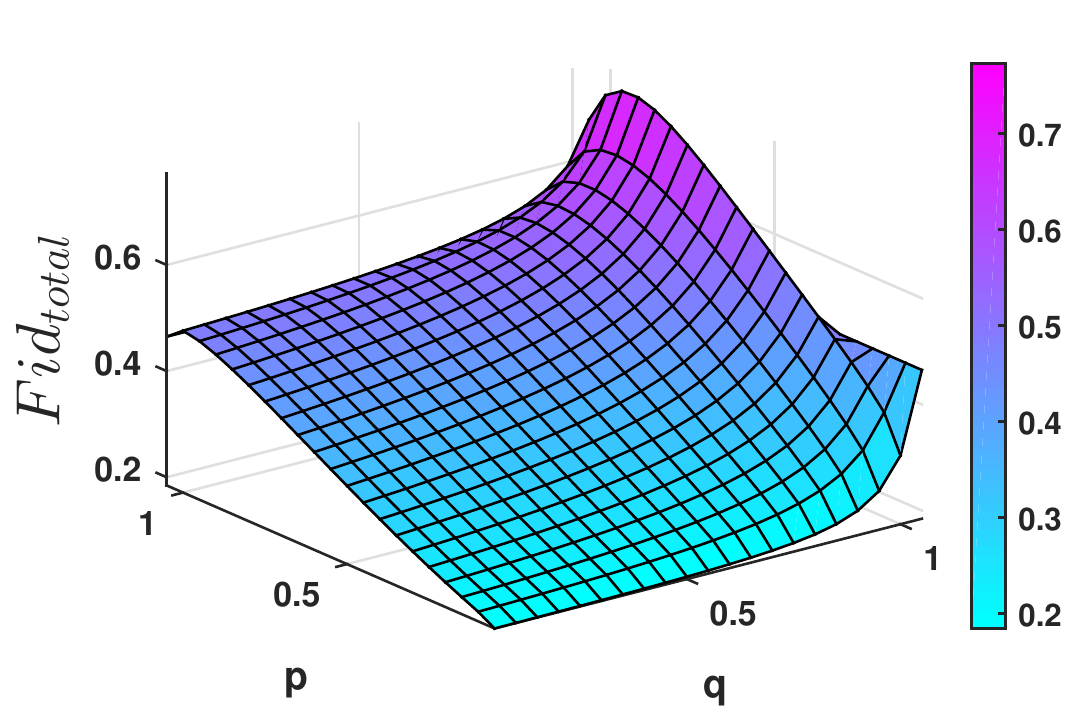}}
        \centering{(c) Pure initial states ($r=0.9$)}
    \end{minipage}
    \hfill
    \begin{minipage}[t]{0.3\textwidth}
        \centering{\includegraphics[width=\textwidth]{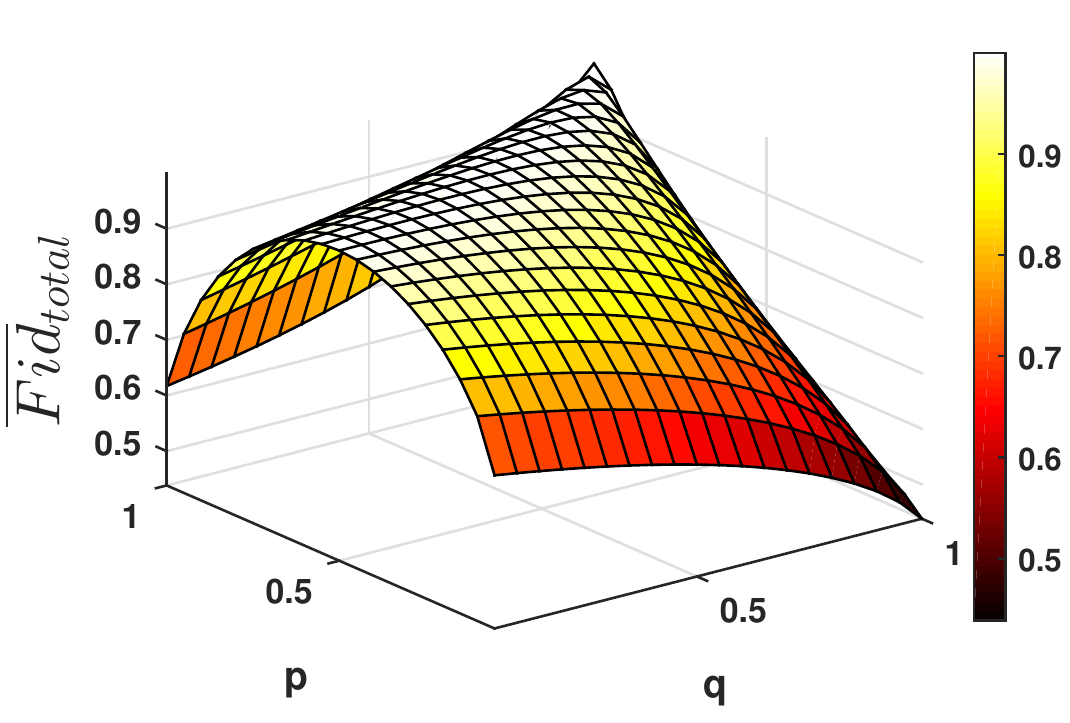}}
        \centering{(d) Mixed initial states ($r=0.1$)}
    \end{minipage}
    \hfill
    \begin{minipage}[t]{0.3\textwidth}
        \centering{\includegraphics[width=\textwidth]{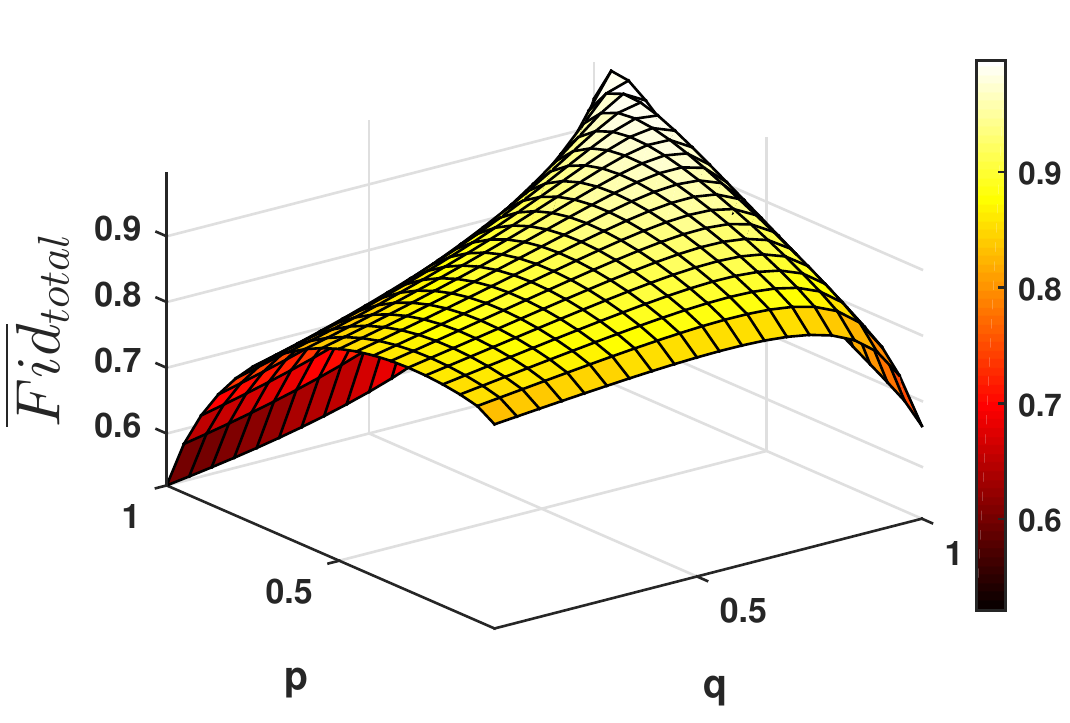}}
        \centering{(e) Mixed initial states ($r=0.6$)}
    \end{minipage}
    \hfill
    \begin{minipage}[t]{0.3\textwidth}
        \centering{\includegraphics[width=\textwidth]{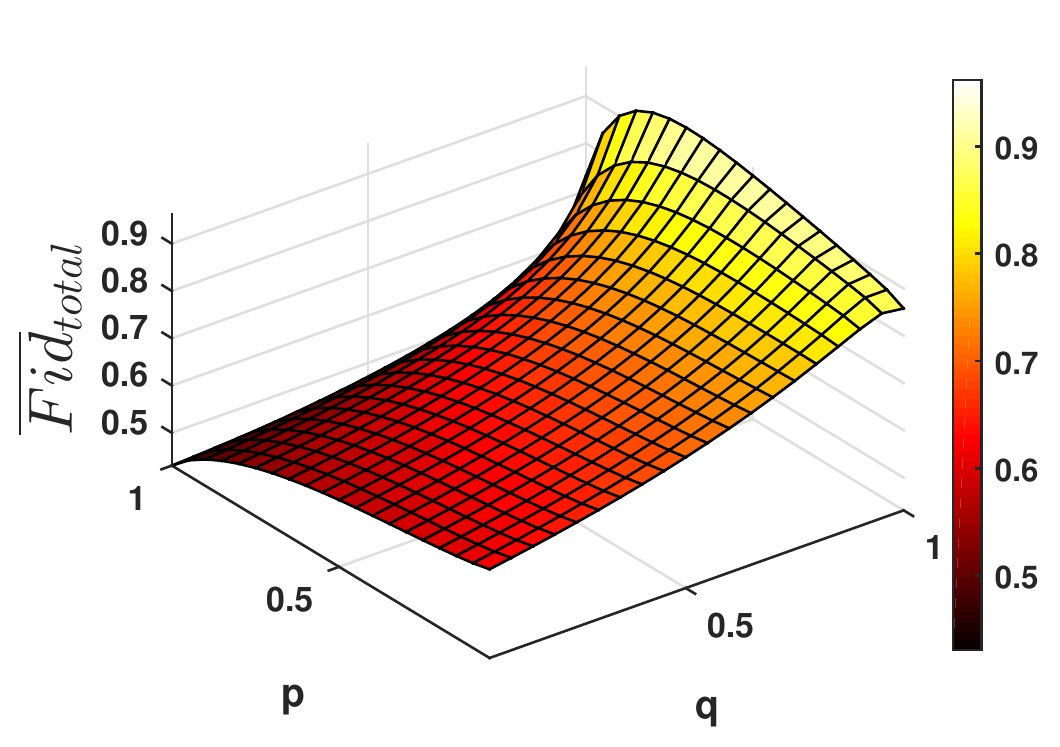}}
        \centering{(f) Mixed initial states ($r=0.9$)}
    \end{minipage}
\caption{ Total fidelity as a function of post-weak measurement strength and pre-weak measurement strength for pure and mixed initial states via Monte-Carlo method for fixed amounts of damping probability $r=0.1,\; 0.6,\; 0.9,$ }\label{fig9}
\end{figure*}

Since the analytical expression for the total final fidelity as a function of $(p,\, q,r)$ in general recovery control is complicated, we use the simulation experiments to study the behavior of the fidelity. The total final fidelity and success probability as a function of post-weak measurement and pre-weak measurement for different amounts of damping probability $r=0.1,\; 0.6,\; 0.9,$ via Monte-Carlo method for pure and mixed initial states are given in Fig. \ref{fig9}.

As Fig. \ref{fig9} shows, to gain higher fidelity one needs to apply stronger post-weak measurement and pre-weak measurement for all amounts of damping probability $r$ and initial state (pure and mixed). Although, the general recovery control fidelity for mixed initial states is higher than the fidelity for pure initial states. For instance, when $r=0.6$, the highest amount of total final fidelity for pure initial states is $Fid_{total} =91.32\% $, where its amount for mixed initial state is $\overline{Fid}_{total} =99.63\% $

\begin{figure}[h!]
    \centering
    \begin{minipage}[t]{0.49\textwidth}
        \centerline{\includegraphics[width=0.85\textwidth]{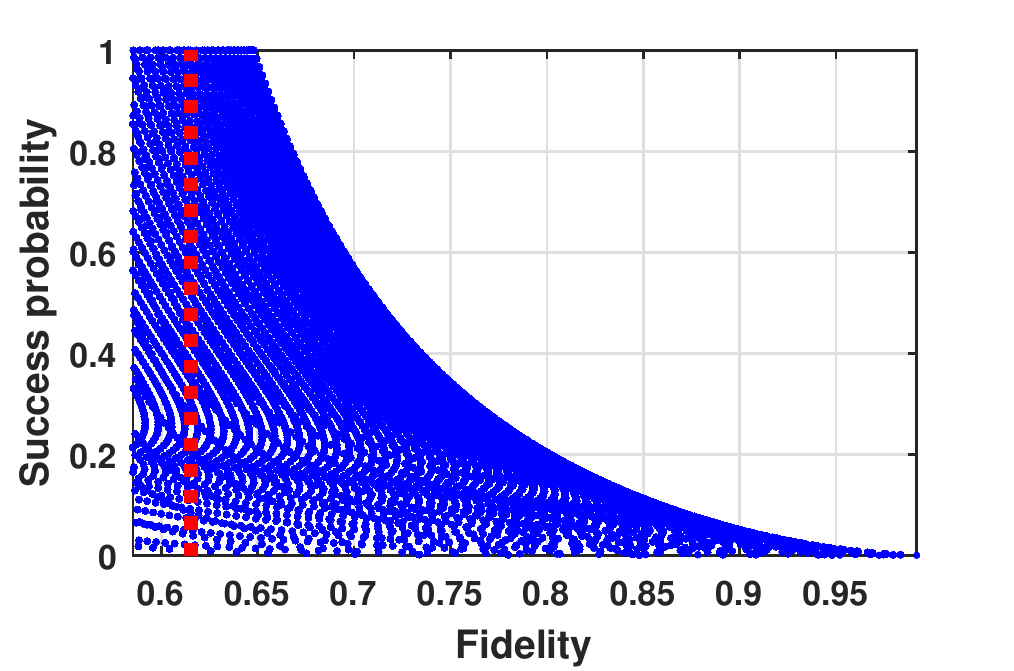}}
        \centerline{(a) Pure initial states}\label{fig10a}
    \end{minipage}
    \hfill
    \begin{minipage}[t]{0.49\textwidth}
        \centerline{\includegraphics[width=0.85\textwidth]{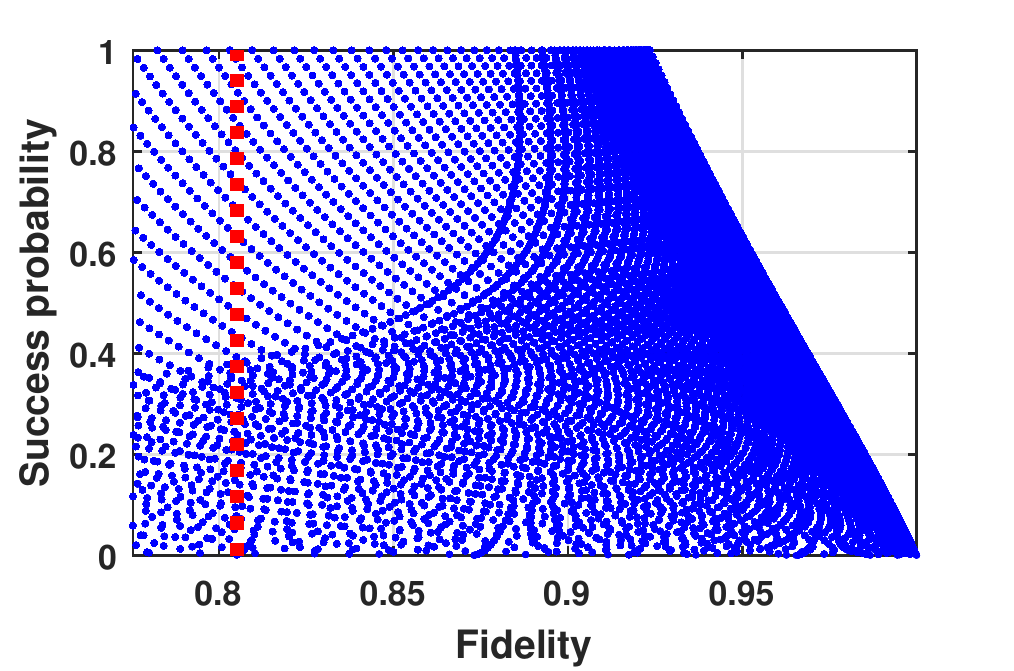}}
        \centerline{(b) Mixed initial states}\label{fig10b}
    \end{minipage}
\caption{ The relation between total fidelity $(Fid_{total} )$ and total success probability $(g_{total} )$ for pure and mixed initial state in general protection scheme via Monte-Carlo method.}\label{fig10}
\end{figure}

In addition, the relation between total fidelity $(Fid_{total})$ and total success probability $(g_{total})$, for all independent real amounts of $(p,\, q)$ from 0 to 1 is given in Fig. \ref{fig10}. In Fig. \ref{fig10} the amount of damping probability is fixed as $r=0.5$ and the relation between total fidelity $Fid_{total} $ and total success probability $g_{total} $ for pure and mixed initial states via Monte-Carlo method is given.

For each amount of fidelity, the maximised success probability is located on the boundary of the diagram. Hence, the optimal fidelities and success probabilities are distributed on the boundary line of the diagram. Each point on the boundary line has a group of measurement strength $(p,\, q)$ amount which we call them as optimal groups of measurement strength. The boundary line of the diagram indicates the relationship between the fidelity and the success probability under the optimal protecting condition.

\section{VII. CONCLUSION} \label{sec7}
In this paper, we proposed a feed-forward control and its reversal to protect the arbitrary initial state of two-qubit. The aim of the feed-forward operation is to make the state of the system robust to the amplitude damping. We consider the recovery in two cases: two-qubit pure initial state and two-qubit mixed initial state. Fidelity and success probability were calculated to evaluate the performance of the control. Theoretical expressions were derived, and specific numerical results were illustrated in plots. We have shown that under complete recovery condition the state of the system can be completely recovered.

\section{ACKNOWLEDGMENTS}

This work was supported by the National Natural Science Foundation of China under grant no. 61573330 and 61720106009

\appendix
\section{Appendix A}
\renewcommand{\theequation}{A\arabic{equation}}
\setcounter{equation}{0}
Here we give the states of the system after the noise channel in case that the result ${\left| 00 \right\rangle} $ corresponding to $M_{00} $ is acquired and after the noise channel just one of the qubits `jump'.

If the first qubit `no jump' and the second qubit `jump', the state of the system becomes:
\begin{equation} \label{a1)}
\begin{aligned}
&{\left| \psi _{M_{00} }^{e_{01} }  \right\rangle}=\\ &\frac{1}{g_{e_{01} } } \left(\beta \sqrt{p} \sqrt{1-p} \sqrt{r} {\left| 00 \right\rangle} +\delta \sqrt{r} \sqrt{1-r} (1-p){\left| 10 \right\rangle} \right)
\end{aligned}
\end{equation}
with probability $g_{e_{01} } =\beta ^{2} p(1-p)r+\delta ^{2} r\, (1-r)\left(1-p\right)^{2} $.

After the noise channel by applying the reversed operation$F_{00} $, the state of the system becomes: ${\left| \psi _{F_{00} }^{e_{01} }  \right\rangle} =F_{00} {\left| \psi ^{e_{01} }  \right\rangle} ={\left| \psi ^{e_{01} }  \right\rangle} $.

At last, after the post-weak measurement the recovered state of the system represented as:
\begin{equation}
\begin{aligned}
{\left| \psi _{N_{00} }^{e_{01} }  \right\rangle} =\frac{1}{g_{e_{01} }^{N_{00} } } \Bigl(&\frac{\beta (1-p)\sqrt{1-p} \sqrt{r} (1-r)}{\sqrt{p} } {\left| 00 \right\rangle} \\&+\frac{\delta \sqrt{r} (1-r)(1-p)\sqrt{1-p} }{\sqrt{p} } {\left| 10 \right\rangle} \Bigr)
\end{aligned}
\end{equation}

with success probability:\\
\begin{align*}
g_{e_{01} }^{N_{00} } =\frac{\beta ^{2} (1-p)^{3} r\, (1-r)^{2} +\delta ^{2} r\, (1-r)^{2} \left(1-p\right)^{3} }{p}.
\end{align*}

In the case that first qubit 'jump' and second qubit `no jump', the state of the system after the noise channel is: ${\left| \psi _{M_{00} }^{e_{10} }  \right\rangle} =\frac{1}{g_{e_{10} } } \left(\gamma \sqrt{p} \sqrt{1-p} \sqrt{r} {\left| 00 \right\rangle} +\delta \sqrt{r} \sqrt{1-r} (1-p){\left| 01 \right\rangle} \right)$
with probability $g_{e_{10} } =\gamma ^{2} p(1-p)r+\delta ^{2} r\, (1-r)\left(1-p\right)^{2} $.

After the feed-forward operation, the state of the system becomes ${\left| \psi _{F_{00} }^{e_{10} }  \right\rangle} =F_{00} {\left| \psi ^{e_{10} }  \right\rangle} ={\left| \psi ^{e_{10} }  \right\rangle} $.

Finally, to recover the state of the system, we apply the post-weak measurement which makes the state of the system as:
\begin{equation}
\begin{aligned}
{\left| \psi _{N_{00} }^{e_{10} }  \right\rangle} =\frac{1}{g_{e_{10} }^{N_{00} } } \Bigl(&\frac{\gamma (1-p)\sqrt{1-p} \sqrt{r} (1-r)}{\sqrt{p} } {\left| 00 \right\rangle} \\&+\frac{\delta \sqrt{r} (1-r)(1-p)\sqrt{1-p} }{\sqrt{p} } {\left| 01 \right\rangle} \Bigr)
\end{aligned}
\end{equation}
with success probability:
\begin{align*}
g_{e_{10} }^{N_{00} } =\frac{\gamma ^{2} (1-p)^{3} r\, (1-r)^{2} +\delta ^{2} r\, (1-r)^{2} \left(1-p\right)^{3} }{p}.
\end{align*}

\bibliographystyle{apsrev4-1} 

\end{document}